\documentclass[final,5p,times,twocolumn]{elsarticle} 

\usepackage{amssymb}
\usepackage[T1]{fontenc}
\usepackage{lineno}

\journal{Nuclear Instruments and Methods in Physics Research A}

\begin{document}
\begin{frontmatter}



\title{Accounting for systematic uncertainties in the Imaging X-ray Polarimetry Explorer (IXPE) detector response}


\author[inst1,inst2]{Stefano Silvestri}
\author[IXPE]{on behalf of the IXPE collaboration}

\affiliation[inst1]{organization={INFN},
            addressline={Largo Bruno Pontecorvo}, 
            city={Pisa},
            postcode={56127}, 
            country={Italy}}

\affiliation[inst2]{organization={Department of Physics, University of Pisa},
            addressline={Largo Bruno Pontecorvo}, 
            city={Pisa},
            postcode={56127}, 
            country={Italy}}

\affiliation[IXPE]{country = {https://ixpe.msfc.nasa.gov/partners\_sci\_team.html}}

\begin{abstract}
Launched on December 9, 2021, the Imaging X-ray Polarimetry Explorer (IXPE) is the first imaging polarimeter ever flown, providing sensitivity in the 2--8 keV range, and during the 2-year initial phase of the mission will sample tens of X-ray sources among different source classes. While most of the measurements will be statistics-limited, for some of the brightest objects observed and long integration times, the systematic uncertainties in the detector response (primarily the effective area, the modulation factor and the absolute energy scale) will be important.
In this contribution, we describe a framework to propagate on high-level observables (e.g.: spectro-polarimetric fit parameters) the systematic uncertainties connected with the response of the detector, that we estimate from relevant ground calibrations and from observations of celestial point sources.
\end{abstract}



\begin{keyword}
Gaseous detectors \sep X-ray astronomy \sep Monte Carlo \sep Calibration
\end{keyword}

\end{frontmatter}


\section{Introduction}
\label{sec:sample1}
IXPE consists of three identical independent telescopes, each comprising a WOLTER-1 type mirror \cite{Wolter52, VanSpeybroeck72} focusing on a Detector Unit (DU) which is a Gas Pixel Detector (GPD \cite{IXPE2021}). X-rays entering the GPD lead to the emission of photo-electrons in the gas mixture where they trigger an avalanche in a direction aligned with that of the electric field of the incident photon, thus providing polarization sensitivity. The GPD is sealed by a beryllium window which has a low Z thus providing excellent transparency to X-rays and preventing gas leakage. The avalanche generated in the gas is drifted by a drift voltage towards a Gas Electron Multiplier (GEM \cite{GEM}) that amplifies the shower which is eventually read-out by an ASIC.
By reconstructing the impact point, the direction of the emitted photo-electron and its energy, IXPE is the first telescope capable of imaging spectro-polarimetric measurements.
We focus on the uncertainty on systematics of the IXPE telescope and analyze their effect on observables using Monte Carlo techniques concentrating on these systematics:
\begin{itemize}
    \item the on-axis effective area, which models the efficiency in the detection of an event as a function of the energy;
    \item the modulation factor (MODF), which models the efficiency of the reconstruction of the photo-electron track;
    \item the modulation response function (MRF), which is the efficiency in detecting a polarized event and is ultimately the product of the first two;
    \item the energy scale, which is influenced by effects that alter the gain of the GEM such as charging.
\end{itemize}

\section{Response functions and perturbation technique}
\begin{figure}
\centering
       \includegraphics[width=0.23\textwidth]{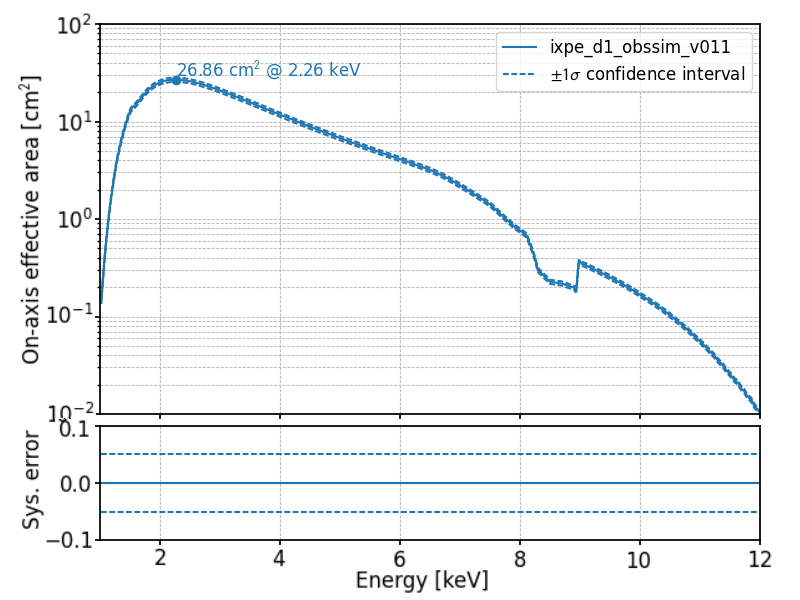}
       \includegraphics[width=0.23\textwidth]{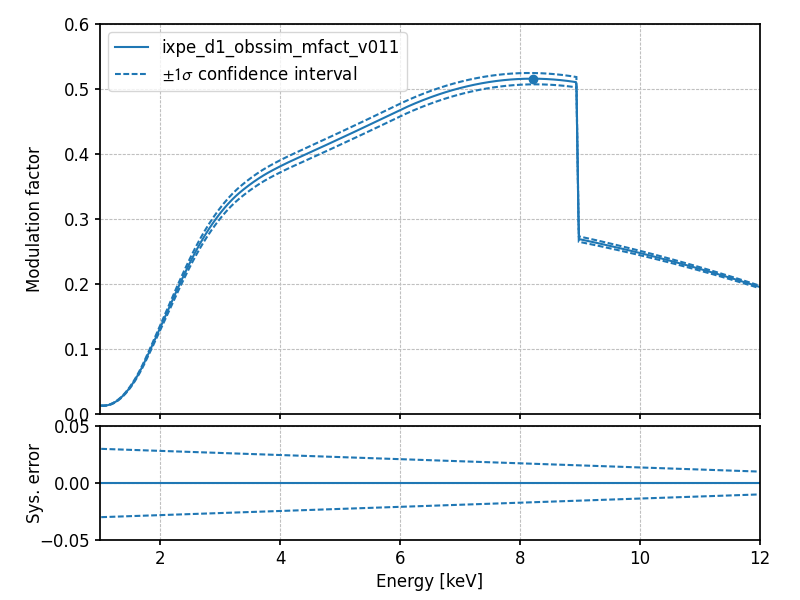}
        \caption{On-axis effective area and modulation factor as functions of the energy with the hypothesized uncertainty on the calibration overlaid. The residual plot shows the relative uncertainty through the energy range.}
 \label{fig:pristine}
 \end{figure}

The response funcions are already provided in the Calibration Database (CALDB) of the mission\footnote{https://heasarc.gsfc.nasa.gov/docs/ixpe/caldb/}, and in this work an uncertainty has been assigned on their shape and value to model systematic errors on the calibration. Figure \ref{fig:pristine} shows the 1$\sigma$ errors that we used for the simulation for the two primary response functions (RF): the on-axis effective area on the left panel and  the MODF on the right for which a trend with the energy has been assumed. Indeed, such a trend is reasonable even considering only the decreasing effect of the spurious modulation with increasing energy, as described in \cite{Rankin22}. Each MRF sample is then just the product of a pair of randomized primaries.

We generate a set of synthetic primary RF as follows: a grid of five points E$_i$ is laid out between 2 and 8 keV, then a random value drawn from a normal distribution with mean value of 1 and a $\sigma$ defined by the energy-dependent relative uncertainty on the RF is assigned at that point $f(E_i)$, finally $f(E_i)$ is smoothed out with a spline and multiplied by the original RF. The MRF is obtained by multiplying the two primaries in order not to lose the information about the combination of fluctuations that produced it.
\begin{figure}
\centering
       \includegraphics[width=0.23\textwidth]{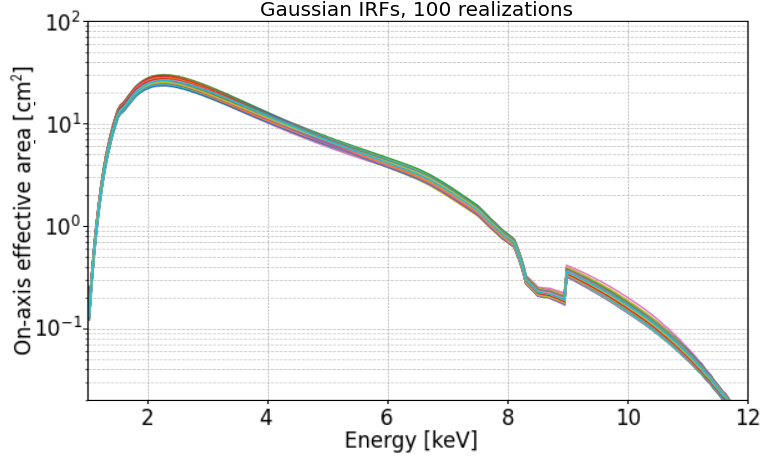}
       \includegraphics[width=0.23\textwidth]{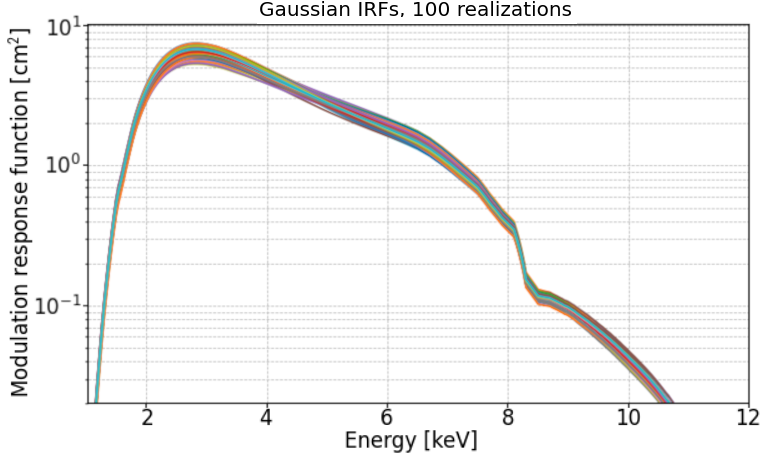}
        \caption{Subset of 100 samples of the synthetic data set of IRFs used for the analysis (only on-axis effective area and MRF are shown). The functions appear smooth and the clearly exhibit variation across the energy band.}
 \label{fig:smeared}
 \end{figure}
A thousand of such synthetic RF are generated to test out the error induced on the parameters estimation with slightly different-than-expected RF. A subset of 100 of such RF, namely the On-axis effective area and the MRF are displayed in figure \ref{fig:smeared}.

\section{Effect of perturbations of the response functions and energy scale}
\begin{figure}[h]
\centering
       \includegraphics[width=0.23\textwidth]{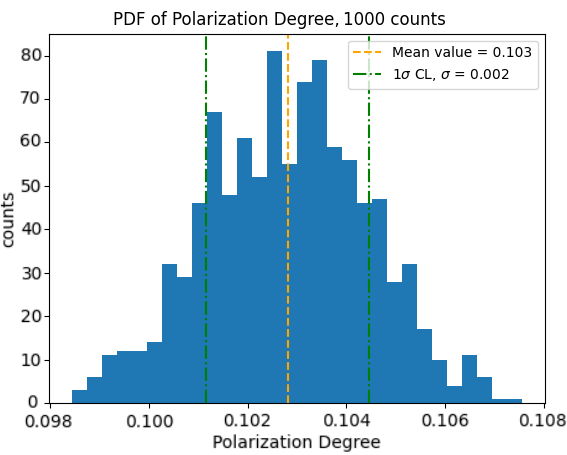}
       \includegraphics[width=0.23\textwidth]{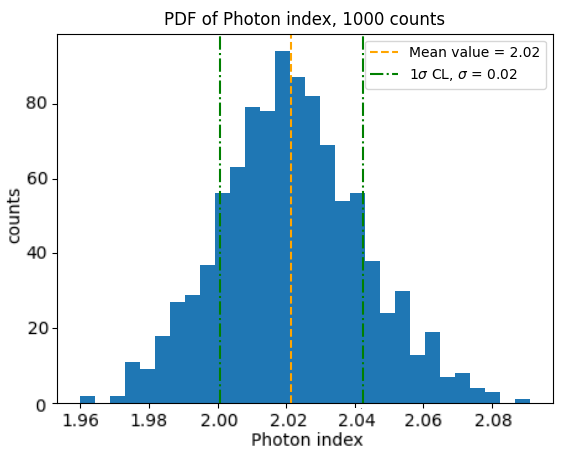}
       \caption{Histograms showing the distribution of the fitted parameters obtained with perturbed IRFs for the Polarization Degree and the Photon Index.}
       \label{fig:histo_irf}
\end{figure}
\begin{figure}[h]
\centering
       \includegraphics[width=0.23\textwidth]{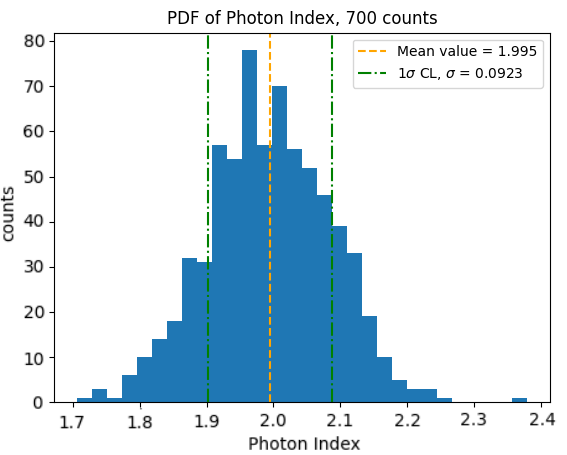}
       \includegraphics[width=0.23\textwidth]{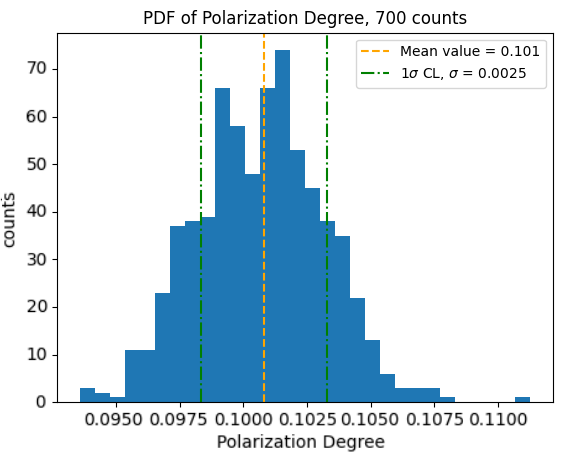}
       \caption{Histograms showing the distribution of the fitted parameters obtained with perturbed energy scale for the Photon Index and the Polarization Degree}
       \label{fig:histo_scale}
\end{figure}

The synthetic set of RF has been used to fold the simulated observation of a point source and the distribution of the parameters inferred with the original RF are shown, highlighting the effect of systematic with respect to the statistical error (Table \ref{tab:mc} and Figure \ref{fig:histo_irf}).
\begin{table}
        \begin{tabular}{c c c c}
            \hline
			Parameter & Target & Result & Systematic IRF\\
			\hline
			Ph. index & 2 & 2.000 $\pm$ 0.001 & 2.02 $\pm$ 0.02\\ 
			Normalization & 10 & 10.00 $\pm$ 0.01 & 10.10 $\pm$ 0.35\\
			Pol. degree & 0.1 & 0.098 $\pm$ 0.0015 & 0.103 $\pm$ 0.002\\
			\hline
		\end{tabular}
		\caption{Simulation target parameters, statistical errors from the fits and outcome of the same fits performed with the Monte Carlo data set of perturbed response functions for Photon index, Normalization and Polarization degree.}
		\label{tab:mc}
\end{table}
The IXPE level 2 data provide Pulse-Invariants (PIs) instead of energy in an OGIP-compliant\footnote{\url{https://heasarc.gsfc.nasa.gov/docs/heasarc/ofwg/docs/spectra/ogip\_92\_007.pdf}} way. PIs are connected to the real photon energy with the energy scale parameter which depends mostly on the gain of the amplification stage. The GEM, however, is not homogeneous and subject to charging: charge build up on the GEM modifying its electric field and altering its gain, which then varies over time and space making this systematic hard to estimate.
The effect of the charging of the GEM has been evaluated by artificially altering the energy scale with a gaussian density with a $\sigma$ of 2$\%$ its original value and analyzing the effects on all spectro-polarimetric parameters, the results are shown in table \ref{tab:PI} and figure \ref{fig:histo_scale}.
\begin{table}
        \begin{tabular}{c c c c}
            \hline
			Parameter & Target & Result & Systematic PI\\
			\hline
			Ph. index & 2 & 2.000 $\pm$ 0.001 & 1.995 $\pm$ 0.09\\ 
			Normalization & 10 & 10.00 $\pm$ 0.01 & 9.95 $\pm$ 0.6\\
			Pol. degree & 0.1 & 0.098 $\pm$ 0.0015 & 0.101 $\pm$ 0.0025\\
			\hline
		\end{tabular}
		\caption{Simulation target parameters, statistical errors for 5$\cdot10^6$ photons and outcome of the same fits performed with the Monte Carlo perturbation of the energy scale for Photon index, Normalization and Polarization degree.}
		\label{tab:PI}
\end{table}

\section{Discussion}
The effect of uncertainties of the order of a few $\%$ on systematics for the IXPE telescope have been estimated. For exceptionally bright sources or long integration time, systematic errors are expected to largely dominate statistical errors for spectral parameters. The polarization degree seems to be far less of a concern even in brighter sources, but in the latter case caution is advised.
Variations on the energy scale due to charging effects also potentially affect the analysis in the very same way. However, charging is a local effect so that dithering or analysis of extended sources are supposed to suffer less from this.

 \bibliographystyle{elsarticle-num-names} 
 \bibliography{Silvestri_Final.bib}





\end{document}